\documentclass[pre,twocolumn,showpacs,preprintnumbers,superscriptaddress,amsmath,amssymb]{revtex4}
\usepackage{graphicx}
\usepackage{subfigure}
\usepackage{mathrsfs}
\usepackage{amsfonts}
\usepackage{amsmath}
\usepackage{appendix}
\usepackage{color}
\usepackage[colorlinks,linkcolor=blue,citecolor=blue]{hyperref}
\usepackage{CJK}
\begin{document}
\begin{CJK}{UTF8}{gbsn}


\title{Quantum-Classical Correspondence Principle for Work Distributions in a Chaotic System}

\author{Long Zhu (朱隆)}
\affiliation{School of Physics, Peking University, Beijing 100871, China}
\author{Zongping Gong (龚宗平ƽ)}
\affiliation{School of Physics, Peking University, Beijing 100871, China}
\author{Biao Wu (吴飙)}
 \affiliation{International Center for Quantum Materials, School of Physics, Peking University, Beijing 100871, China}
 \affiliation{Collaborative Innovation Center of Quantum Matter, Beijing 100871, China}
 \affiliation{Wilczek Quantum Center, College of Science, Zhejiang University of Technology, Hangzhou 310014, China}
 \affiliation{Synergetic Innovation Center for Quantum Effects and Applications (SICQEA), Hunan Normal University, Changsha 410081, China}
\author{H. T. Quan(ȫ全海涛)}
\email{htquan@pku.edu.cn}
\affiliation{School of Physics, Peking University, Beijing 100871, China}
\affiliation{Collaborative Innovation Center of Quantum Matter, Beijing 100871, China}
\date{\today}
\pacs{05.70.Ln, 05.30.-d, 05.45.Pq}

\begin{abstract}
We numerically study the work distributions in a chaotic system and examine 
the relationship between quantum work and classical work. Our numerical results 
suggest that there exists a correspondence principle between quantum and classical 
work distributions in a chaotic system. This correspondence was proved 
for one-dimensional (1D) integrable systems in a recent 
work [Jarzynski, Quan, and Rahav, Phys. Rev. X 5, 031038 (2015)]. Our investigation 
further justifies the definition of quantum work via two point energy measurements.
\end{abstract}
\maketitle

\section{\label{sec:level1}Introduction}

In the past two decades, substantial developments have been made in the field of nonequilibrium statistical mechanics in small systems \cite{Jarzynski2011,Seifert2012}. A set of exact relations of fluctuations regarding work \cite{Jarzynski1997a,Jarzynski1997b,Crooks1998,Crooks1999,Crooks2000,Gong2015}, heat \cite{Park2014,Gong2015}, and entropy production \cite{Seifert2005} have been discovered. They are now collectively known as Fluctuation Theorems (FT) \cite{Jarzynski1997a,Jarzynski2011,Seifert2012,Gong2015}. These theorems are valid in 
processes that are arbitrarily far from equilibrium, and have significantly advanced our understanding about the physics of nonequilibrium processes in small systems. These FT not only imply the second law of thermodynamics but also predict quantitatively the probabilities of the events, which ``violate" the second law in small systems. Despite these developments, there are still some aspects of these FT that have not been fully understood. One of these aspects is the definition of quantum work. For an isolated quantum system, there are various definitions of quantum work \cite{Talkner2015a}. However, it is found that within a large class of definitions, only one \cite{Talkner2014,Talkner2015} of them satisfies the FT. That is, the work defined through two point energy measurements: one at the beginning and the other at the end of a driving process \cite{Kurchan2000,Tasaki2000}. This definition of quantum work, though leading to FT, seems ad hoc because the collapse of the wave function \cite{Neumann1955}, which plays a central role in determining the work \cite{Talkner2015}, brings profound interpretational difficulty to the definition of quantum work \cite{Jarzynski2015}. It is thus important to justify the definition of quantum work, i.e., to find other independent evidences (besides the validity of FT) to support the definition of quantum work via two point energy measurements. Since the correspondence principle \cite{Bokulich2008} is a bridge connecting quantum and classical mechanics, we believe that the correspondence principle for work distributions (if there is one) can be a good evidence to justify this definition.

Recently, the relationship between quantum and classical work distributions in 
one dimensional (1D) integrable systems has been carefully studied \cite{Jarzynski2015}. By employing the semiclassical method \cite{Delos1986,Littlejohn1992}, it is rigorously proved that such a correspondence principle exists for work distributions in 1D integrable systems when the quantum work is defined via two point energy measurements. 
Nevertheless, for a generic system, especially a chaotic one, the correspondence principle for work distributions has not been explored so far. 
In this article, we try to address this issue following the efforts of Ref. \cite{Jarzynski2015}. If the correspondence principle for work distributions also exists in chaotic systems, the justification of the definition of quantum work via two point energy measurements can be extended to chaotic systems.

Among various chaotic systems, one of the most extensively studied systems is the billiard systems \cite{Gutzwiller1990,Stone2005}. In this article, we numerically study the work distribution in a driven billiard system---a ripple billiard \cite{Wubiao2002} with moving boundaries. We numerically compute the time evolution of the chaotic billiard system in both quantum and classical regimes, and then study the relationship between the distributions of quantum work and classical work. Our numerical results suggest 
that the correspondence principle applies in this context.

The paper is organized as follows. In Sec. II we introduce the quantum and the classical transition probabilities, which are used in the calculation of the work distributions. In Sec. III, we introduce the billiard model. In Sec. IV, we present our numerical results and our analysis. In Sec. V, we make some concluding remarks. The numerical method is presented in the Appendix.

\section{\label{sec:level1}Classical and Quantum Transition Probabilities}
We consider a quantum system, which is driven in a nonequilibrium process 
from time $t=0$ to $t=\tau$. This is usually characterized by the work parameter
of the system $b$ that changes from $A$ to $B$. The work parameter $b$ can
be  the position of a piston or the spring constant of a harmonic oscillator or else. 
For  this nonequilibrium process, its  work distribution function  can be expressed as \cite{Talkner2007,Jarzynski2015}:
\begin{equation}
P^{Q}(W)=\sum_{m,n}P^{Q}_{A}(m)P^{Q}(n^{B}|m^{A})\delta(W-E_{n}^{B}+E_{m}^{A}).
\label{quantumwork}
\end{equation}
It is clear that the work distribution function is determined by two factors.
The first one is the initial thermal distribution function $P^{Q}_{A}(m)=e^{-\beta E_{m}^{A}}/Z^Q_{A}$, $Z^Q_{A}=\sum_{m}e^{-\beta E_{m}^{A}}$, where $E^A_m$ is the energy of the $m$th eigenstate when $b=A$, $Z^Q_A\equiv\sum_m e^{-\beta E^A_m}$ is the partition function and $\beta$ is the inverse temperature of the initial thermal state. The second one is the transition probability $P^{Q}(n^B | m^A)$ 
between the initial and the final energy eigenstates 
during the driving process. Similarly, the work distribution function of  the corresponding 
classical system is \cite{Jarzynski2015}
\begin{equation}
P^{C}(W) \approx \sum_{m,n}P^{C}_{A}(m)P^{C}(n^{B}|m^{A})\delta(W-E_{n}^{B}+E_{m}^{A}),
\label{classicalwork}
\end{equation}
where $P^{C}_{A}(m)$ ($P^C(n^B|m^A)$) is the classical counterpart of $P^{Q}_{A}(m)$ ($P^Q(n^B|m^A)$). 

The transition probability of the quantum system is defined as
\begin{equation}
P^Q(n^B| m^A)\equiv \left| \left\langle n^B\right| \hat{U}(t) \left| m^A \right\rangle \right|^2,
\label{pq}
\end{equation}
where $\left| m^A \right\rangle$ and $\left| n^B \right \rangle$, respectively, represent the $m$th eigenstate at the initial time $t=0$ and the  $n$th eigenstate at the final time $t=\tau$. Accordingly, $P^Q(n^B|m^A)$ denotes the quantum transition probability from the $m$th eigenstate when $b=A$, to the $n$th eigenstate when $b=B$. $\hat{U}(t)$ is the unitary operator satisfying $i\hbar \partial \hat{U}(t)/\partial t=\hat{H}(t)\hat{U}(t)$, where $\hat H(t)$ is the time-dependent Hamiltonian of the system.

While the definition of the quantum transition probability is straightforward, the definition of its classical counterpart is a bit subtle. 
The classical transition probability $P^C(n^B|m^A)$ is defined as follows \cite{Jarzynski2015,Schwieters1995b}. 
Initially the microscopic states are evenly sampled from the energy shell $E=E_m^A$ in the classical phase space (see Fig. \ref{fig1}). 
Each microscopic state is represented by a phase-space point. The initial states then undergo Newtonian dynamics governed by $H(t)$, 
when the work parameter $b(t)$ is varied according to a given protocol. 
The corresponding classical transition probability is defined as
\begin{equation}
P^C(n^B | m^A)\equiv \frac{N_{in}}{N_{total}},
\label{eq_pro}
\end{equation}
where $N_{in}$ is the number of representative points which fall into the energy window 
$(E_n^B,E_{n+1}^B)$ at $t=\tau$, and $N_{total}$ is the total number of 
the representative points. So, $P^C(n^B| m^A)$ is the probability of a classical particle 
whose energy is initially $E_m^A$  and finally falls into the energy 
window $(E_n^B,E_{n+1}^B)$ at $t=\tau$.

In order to study the initial thermal distribution $P_A^C(m)$ and $P_A^Q(m)$, we need to clarify the density of states. The density of states for the classical system is given by 
\begin{equation}
\bar{\rho}(E)=\int \frac{d^{d}q d^{d}p}{(2\pi \hbar)^{d}} \delta(E-H(p,q)),
\end{equation}
where $q$ and $p$ are the coordinate and the momentum of the system, $d$ is the spatial dimension, $H(p,q)=H(0)$ is the initial Hamiltonian. 
Accordingly, the initial thermal distribution for this classical system reads 
\begin{equation}
P^C_{A}(m)= \int_{E_{m}^{A}}^{E_{m+1}^{A}}  \bar{\rho}(E) \frac{1}{Z_{A}^{C}} e^{-\beta E} dE ,
\end{equation}
where
\begin{equation}
Z_{A}^{C}=\int \frac{d^{d}q d^{d}p}{(2\pi \hbar)^{d}} e^{-\beta H(p,q)}
\end{equation}
is the classical partition function.

For the quantum system, according to Gutzwiller \cite{Gutzwiller1990} the semiclassical density of states is equal to the summation of the classical density of states $\bar{\rho}(E)$ and an oscillating correction term $\tilde{\rho}(E)$ \cite{Gutzwiller1971,Berry1983,Gutzwiller1990,Bokulich2008,Engl2015}
\begin{equation}
\rho(E)=\bar{\rho}(E)+ \tilde{\rho}(E).
\end{equation}
The oscillation part $\tilde{\rho}(E)$ has an origin in the classical period orbits, and is absent for classical systems. To the first order approximation, or on an energy scale larger than the period of $\tilde{\rho}(E)$, we can ignore the oscillation part $\tilde{\rho}(E)$ and  keep only the average density of states $\bar{\rho}(E)$. Therefore, the initial thermal distribution for the quantum system is approximately equal to its classical counterpart 
\begin{equation}
\begin{split}
P^Q_{A}(m)&= \int_{E_{m}^{A}}^{E_{m+1}^{A}}  \rho(E) \frac{1}{Z_{A}^{Q}} e^{-\beta E} dE \\
&\approx \int_{E_{m}^{A}}^{E_{m+1}^{A}}  \bar{\rho}(E) \frac{1}{Z_{A}^{C}} e^{-\beta E} dE = P^C_{A}(m).\label{appro}
\end{split}
\end{equation}
As a result, the comparison between the quantum (\ref{quantumwork}) and the classical (\ref{classicalwork}) work distributions is reduced to the comparison between the quantum $P^Q(n^B | m^A)$ (\ref{pq}) and the classical $P^C(n^B | m^A)$ (\ref{eq_pro}) transition probabilities. 


Both the quantum transition probability (\ref{pq}) and its classical counterpart (\ref{eq_pro}) 
are computable. However, unlike the integrable systems \cite{Jarzynski2015}, 
for chaotic systems,  we have to resort to the numerical method as the analytical semiclassical (WKB) wave function \cite{Griffiths1995} of the energy eigenstate in a fully chaotic system is usually unavailable \cite{Berry1977b,Berry1979,Berry1983,Ott2002,Stone2005,Bokulich2008}. 
We will introduce the model in the next section. Our numerical results will be presented in the Sec. IV.



\section{\label{sec:level1}The Ripple Billiard Model}
A prototype model widely studied in quantum chaos is a static billiard system 
whose boundaries are fixed and the Hamiltonian is time-independent.  
For our study,  we choose a ripple billiard \cite{Wubiao2002}  whose sinusoidal 
boundaries move in opposite directions. 
The advantage of choosing the ripple billiard instead of other more extensively studied systems in literature, such as the stadium billiard \cite{Gutzwiller1990,Wisniacki2003}, is that each entry of its Hamiltonian matrix can be expressed in terms of elementary functions \cite{Wubiao2002}. Thus the eigenenergies and eigenstates of the system at any moment of time can be obtained through exact numerical diagonalization, which is usually not doable in chaotic systems. 
As a result, we can accurately simulate  the quantum  dynamical evolution in the chaotic system, 
which is usually a big challenge \cite{challenge}. 
In addition, the degree of chaoticity of the model can be controlled by tuning the geometric parameters $a$, $b$ and $L$, which enables us to study the influence of the degree of chaoticity easily.


\begin{figure}[tbh]
\centering
\subfigure[]{
\label{fig.sub.1}
\includegraphics[width=9cm, angle=0]{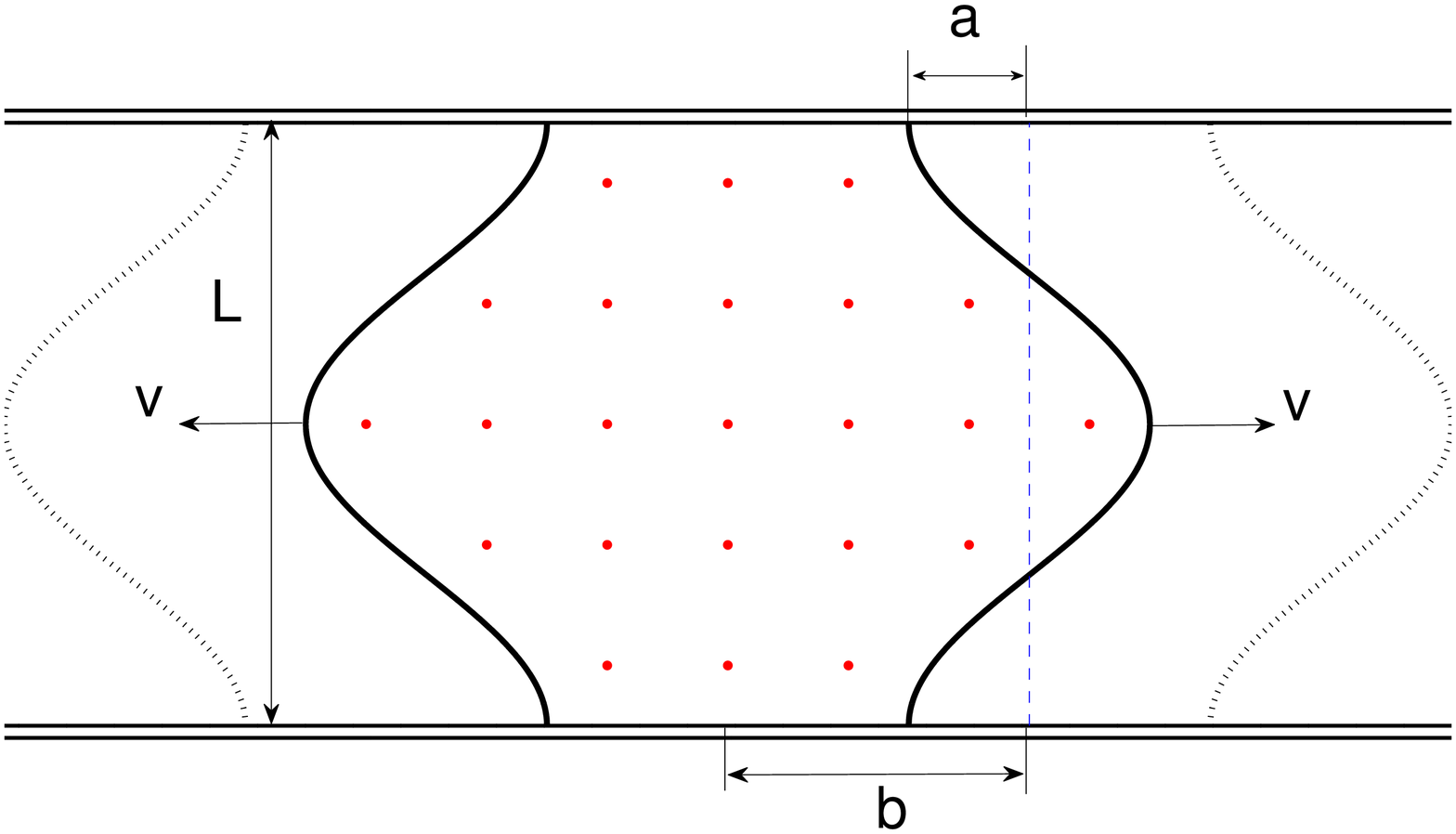}}
\subfigure[]{
\label{fig.sub.2}
\includegraphics[width=7cm,angle=0]{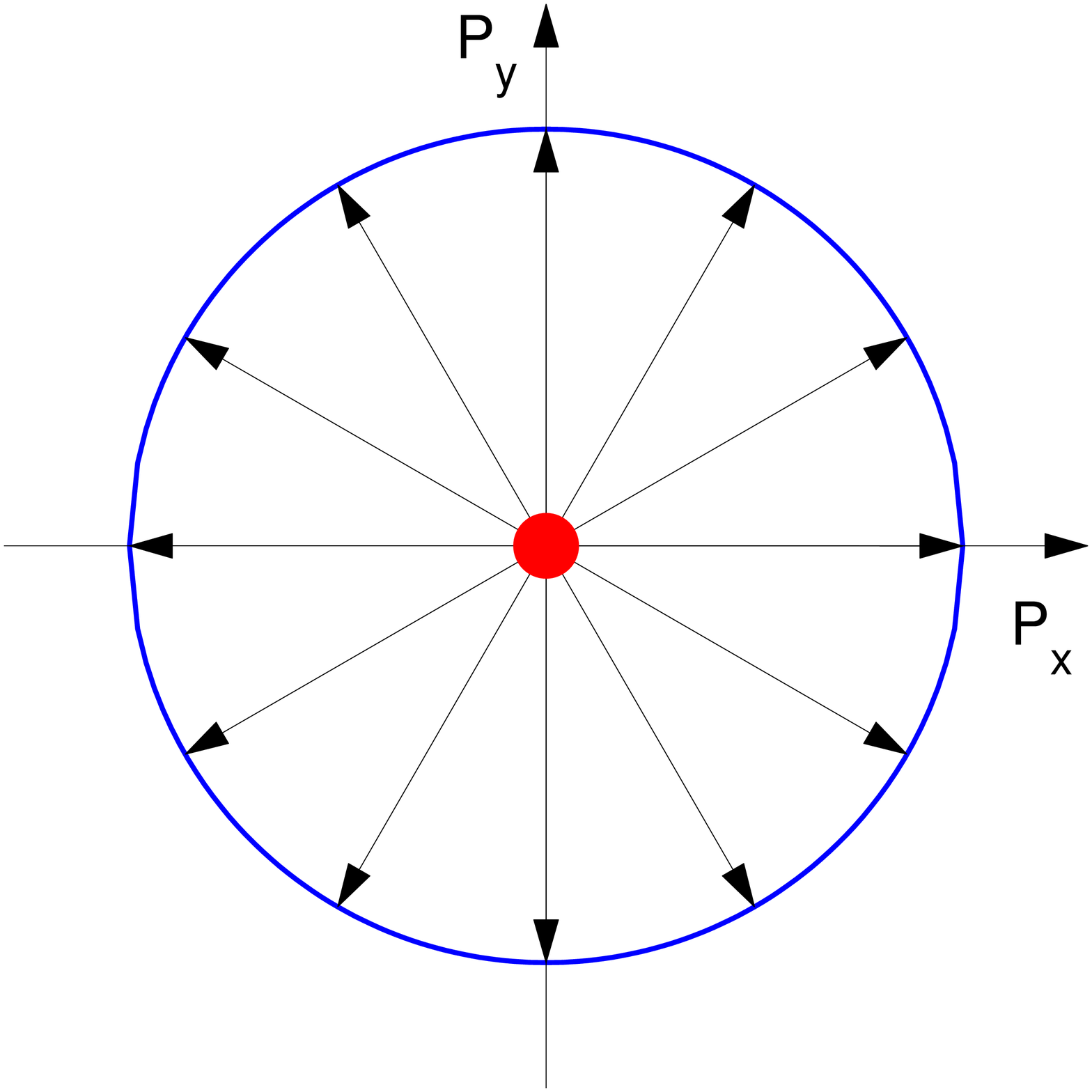}}
\caption{Ripple billiard with moving boundaries. The coordinates in the position space and the momentum space are denoted by $(x,y)$ and $(p_{x},p_{y})$, respectively. Parameter $b$ is varied in time $b(t)=b_0+vt$, where $b_0=A$. The parameter $a$ represents the ripple amplitude. The red dots in the Fig.~\ref{fig.sub.1} show that the initial states are evenly sampled in the coordinate space. Fig.~\ref{fig.sub.2} shows that the initial states are evenly sampled in the momentum space from an energy shell $E_{m}^A=(p_{x}^{2}+p_{y}^{2})/2M$, where $M$ is the mass of the billiard ball. }
 \label{fig1}
\end{figure}

The ripple billiard \cite{Wubiao2002} with both boundaries moving at the same speed 
and in the opposite direction is illustrated in Fig.~\ref{fig1}. In our model, the work parameter is 
the (half) length $b$ of the billiard. We move both the curved boundaries because 
the symmetry can help us simplify the calculation. The position and the momentum of the particle 
inside the billiard  are denoted by $(x, y)$ and $(p_{x},p_{y})$, respectively.  
This model system can be characterized by the following parameters: 
$b_0,a, L$, $v$, and $\tau$. $b_0$ is the initial length, $a$ and $L$ are the parameters characterizing the 
boundary shape of the ripple billiard, 
while $v$ is the speed of the moving boundary and $\tau$ is the total driving time. 
In an appropriately chosen coordinate, the boundaries of the ripple billiard can be depicted by
\begin{equation}
f(y,t)=\pm[b(t)-a\cos(2\pi y/L)],
\end{equation}
where $a$ represents the ripple amplitude and $b(t)=b_0+vt$ denotes the length at time $t$. 
When $a$ is decreased, the system becomes ``less chaotic". When $a=0$, the system becomes integrable. Since we are interested in the dynamical evolution of a chaotic system, we fix $a$ at a finite number and vary $b$ in time. For a large $a$ the system is always in the deep chaotic regime.

This kind of driven quantum systems are of interest in the field of mesoscopic physics and have been studied by Cohen {\it et al.}~\cite{Cohen2002}. 
For 1D system, work distributions in a quantum and a classical billiard have been obtained in Refs.~\cite{Quan2012,Lua2005}. For 2D systems, some brief results regarding time-dependent integrable quantum billiards (rectangular and elliptical billiards) have been reported by Shmelcher {\it et al.} \cite{Schmelcher2009}, but no results about chaotic billiards were reported there. 

It should be emphasized that the counterpart of the energy eigenstate $\left|m^{A} \right\rangle$ in the classical system is a microcanonical ensemble, namely the collection of representative points evenly sampled from a 3D ``energy shell" $E=E_{m}^{A}$ in a 4D phase space. In the coordinate space, the position components of the representative points $(x,y)$ are evenly sampled inside the potential well (see Fig.~\ref{fig1}(a)). In the momentum space, the momentum components $(p_{x},p_{y})$ are evenly sampled from the energy shell $E_{m}^{A}=(p_{x}^{2}+p_{y}^{2})/2M$ (see Fig.~\ref{fig1}(b)). Such a choice of sampling assures a uniform (isotropic) distribution in the coordinate (momentum) space after the local average over a vicinity which is small compared to the size of the potential well but large compared to the quantum wave length 
\cite{Berry1977b,Kaufman1988,Ott2002}. For simplicity, we set the mass of the billiard ball to be $M=0.5$. In the next section, we numerically simulate the classical and the quantum evolution of the driven ripple billiard system and compare these two transition probabilities to check if there exists the correspondence principle between the transition probabilities in this chaotic system.

\section{\label{sec:level1}Numerical Results}
We develop a method to accurately calculate the quantum transition probabilities 
in a ripple billiard system with moving boundaries. Technical details can be found 
in the Appendix. We only present the numerical results in the main text. 

\begin{figure}[tbh]
\includegraphics[width=9cm, clip]{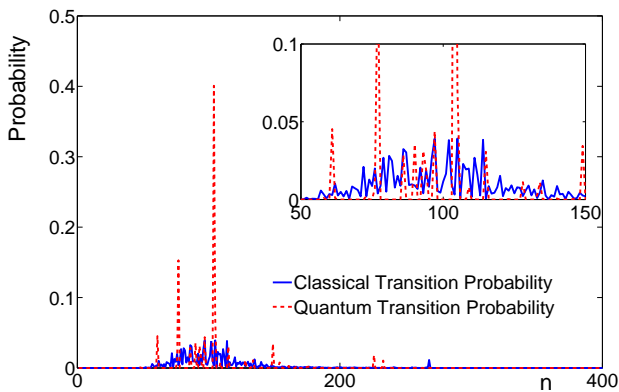}
\caption{ Comparison between quantum $P^Q(n^B | m^A)$ and classical $P^C(n^B | m^A)$ transition probabilities. Here the horizontal axis labels quantum number $n$. 
The parameters are set as $b_0=0.5, a=0.2, L=1.0, v=2.0, \tau=0.4, \hbar=1.0$. The initial state is set to be the 100th eigenstate, namely $m=100$. The partial enlarged details are shown in the insetting. }
\label{fig2}
\end{figure}

\begin{figure*}[tbh]
\includegraphics[width=18cm, angle=0]{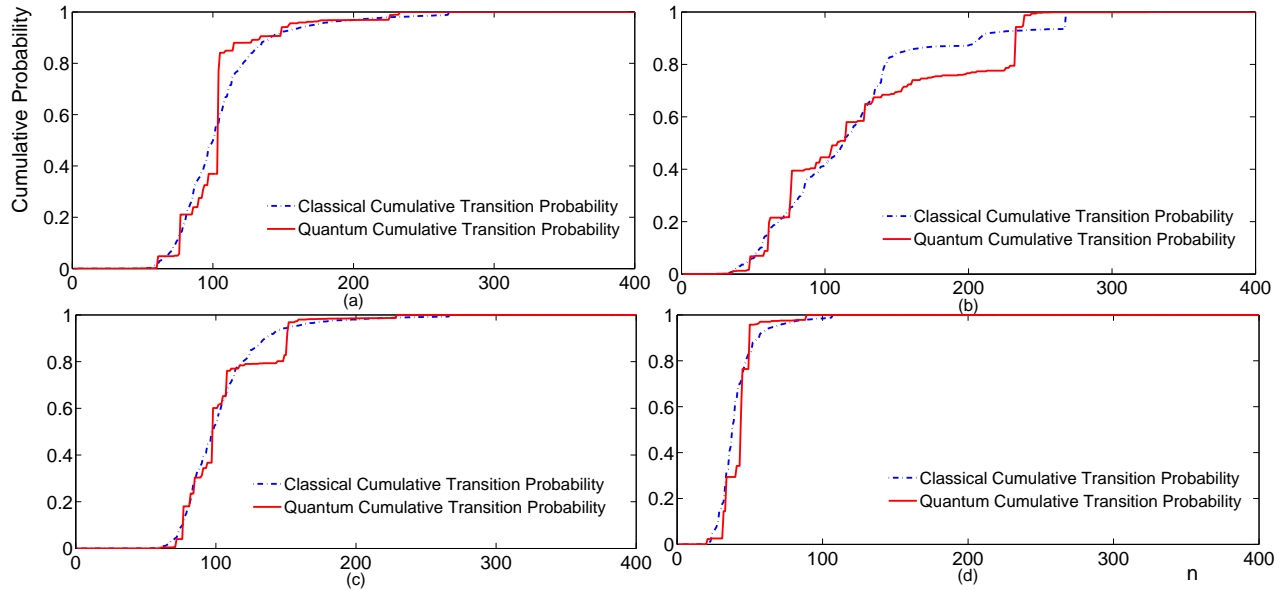}
\caption{ Comparison between quantum $S_n^{Q}$ and classical $S_n^{C}$ cumulative transition probabilities with various sets of parameters $a$, $v$ and $\hbar$. Here the horizontal axis labels quantum number n and the vertical axis labels the cumulative transition probability $S_n^{Q}(S_n^{C})$. 
The parameters in the four figures are respectively set as: (a) $b_0=0.5, a=0.2, L=1.0, v=2.0, \tau=0.4, \hbar=1.0$; (b) $b_0=0.5, a=0.2, L=1.0, v=8.0, \tau=0.1, \hbar=1.0$; (c) $b_0=0.5, a=0.3, L=1.0, v=2.0, \tau=0.4, \hbar=1.0$; (d) $b_0=0.5, a=0.2, L=1.0, v=2.0, \tau=0.4,\hbar=1.5571$.}
\label{fig3}
\end{figure*}

An example of the comparison between the quantum and the classical transition probabilities 
is shown in Fig.~\ref{fig2}. The parameters are  $b_0=0.5,a=0.2,L=1.0,v=2.0,\tau=0.4$, and the Planck's constant $\hbar=1.0$. The initial state is  the 100th eigenstate ($m=100$). In Fig.~\ref{fig2} we notice that (i) both the quantum and the classical transition probabilities are less regular than those in the 1D integrable case \cite{Jarzynski2015} and (ii) the quantum transition probabilities are sparse and discrete, while the classical transition probabilities are quasi-continuous and spread in a wide range of the energy spectrum. In order to compare these two transition probabilities in a better way, we plot the {\it {cumulative transition probabilities}} in Fig.~\ref{fig3} for different sets of parameters. From Fig.~\ref{fig3} we can see that the quantum and the classical transition probabilities do not collapse onto the same curve, but are very close to each other. Especially
the quantum cumulative transition probability curve oscillates around the smooth classical cumulative transition probability curve. This phenomenon is reminiscent of the results in 1D integrable systems \cite{Jarzynski2015}, where it 
has been explained as a consequence of the interference of different classical trajectories. Although we cannot clearly see the correspondence between the quantum and the classical transition probabilities in Fig.~\ref{fig2}, we obviously observe the convergence in Fig.~\ref{fig3}, which implies a corresponding principle between the transition probabilities. 

Having demonstrated the correspondence between the quantum and the classical transition 
probabilities, we will study the effects of the degree of chaoticity (characterized by $a$), 
the speed ($v$) of the moving boundary, and the value of the Planck constant ($\hbar$) on the convergence of the two transition probabilities. To this end, we introduce a measure to quantify the distance between the classical and the quantum cumulative transition probabilities. 
The measure we choose is the root-mean-square error (RMSE) (see textbooks on mathematical statistics, for example Ref.~\cite{book}).
At time $t$, the RMSE $R(t)$ between the two cumulative transition probabilities is defined as
\begin{equation}
R(t)\equiv \sqrt{\frac{\sum_{n;S_n^C\neq S_n^Q}(S_n^C(t)-S_n^Q(t))^2}{N(t)}},
\label{rmse}
\end{equation}
where $S_n^C(t)=\sum\limits_{k=1}^{n}P^C(k^{B(t)}| m^A)$ is the classical cumulative transition probability at time $t$ and $S_n^Q(t)=\sum\limits_{k=1}^{n}P^Q(k^{B(t)}| m^A)$ is its quantum counterpart. We use $k^{B(t)}$ instead of $k^B$ to emphasize that the work parameter $B$ is time-dependent. The sum of the squared difference between $S_n^C(t)$ and $S_n^Q(t)$ is taken over all $n$ where these two quantities are not equal. $N(t)$ is the total number of the quantum numbers at which these two cumulative transition probabilities are different. Roughly speaking, $R(t)$ is the average of the local deviations between these two cumulative transition probabilities at time $t$. $R(t)=0$ means that the two distributions are identical. The larger $R(t)$ is, the more distinct the two probabilities $S_n^C(t)$ and $S_n^Q(t)$ are. We note that the correspondence principle implies the convergence between the classical and the quantum transition probabilities in some average sense \cite{Jarzynski2015}. The RMSE quantifies the average distance between two probability distributions. Hence, we believe that the RMSE can be a good measure to quantify the applicability of the correspondence principle.

\begin{figure}[tbh]
\includegraphics[width=9cm, angle=0]{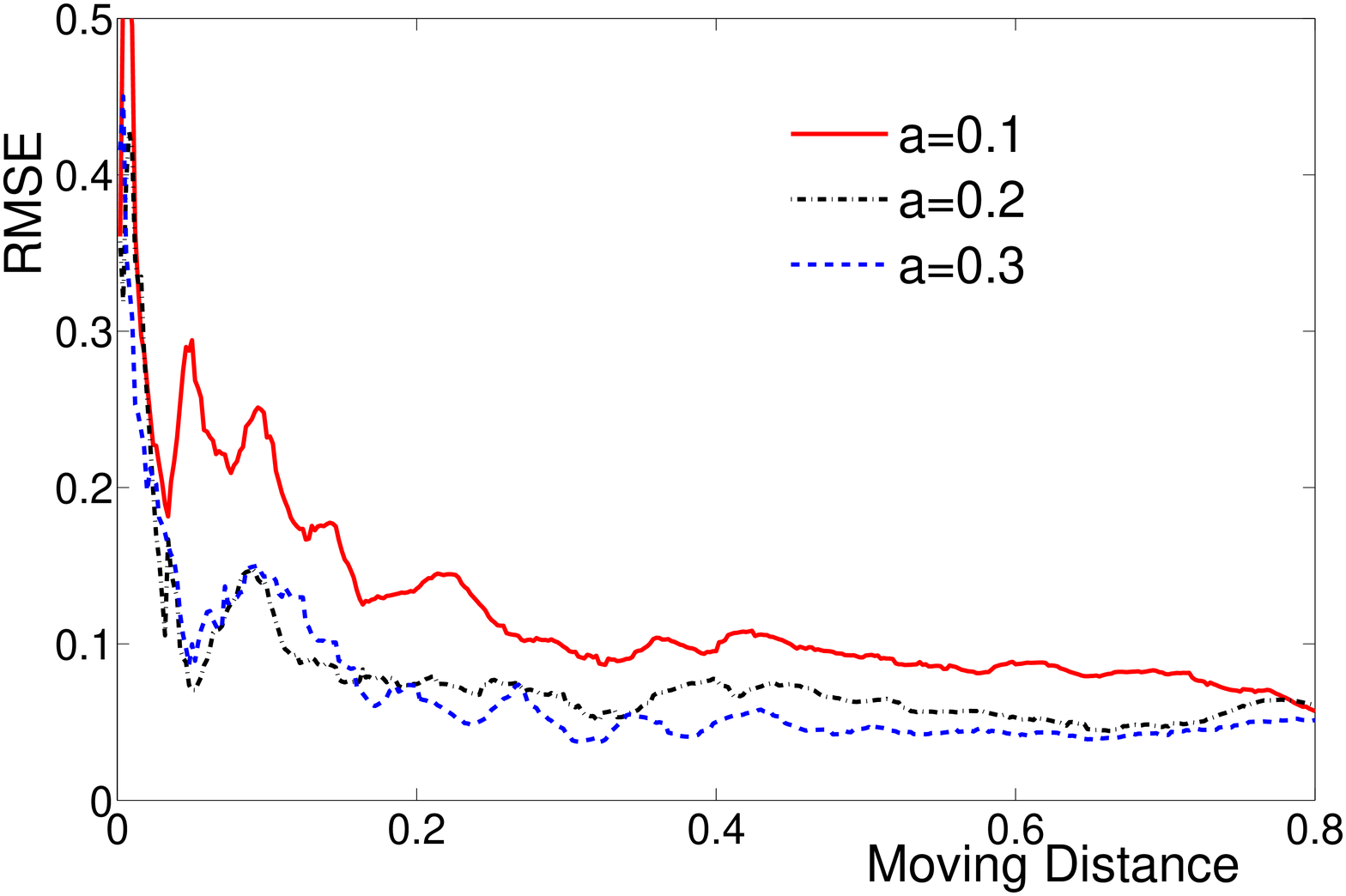}
\caption{ Comparison of RMSE as a function of boundary-moving distance (one side) at different degrees of chaoticity $a=0.1,0.2,0.3$. The other parameters are set as $b_0=0.5, L=1.0, v=2, \tau=0.4, \hbar=1.0$ and the initial state is the 100th eigenstate ($m=100$).}
\label{fig4}
\end{figure}

\begin{figure}[tbh]
\includegraphics[width=9cm, angle=0]{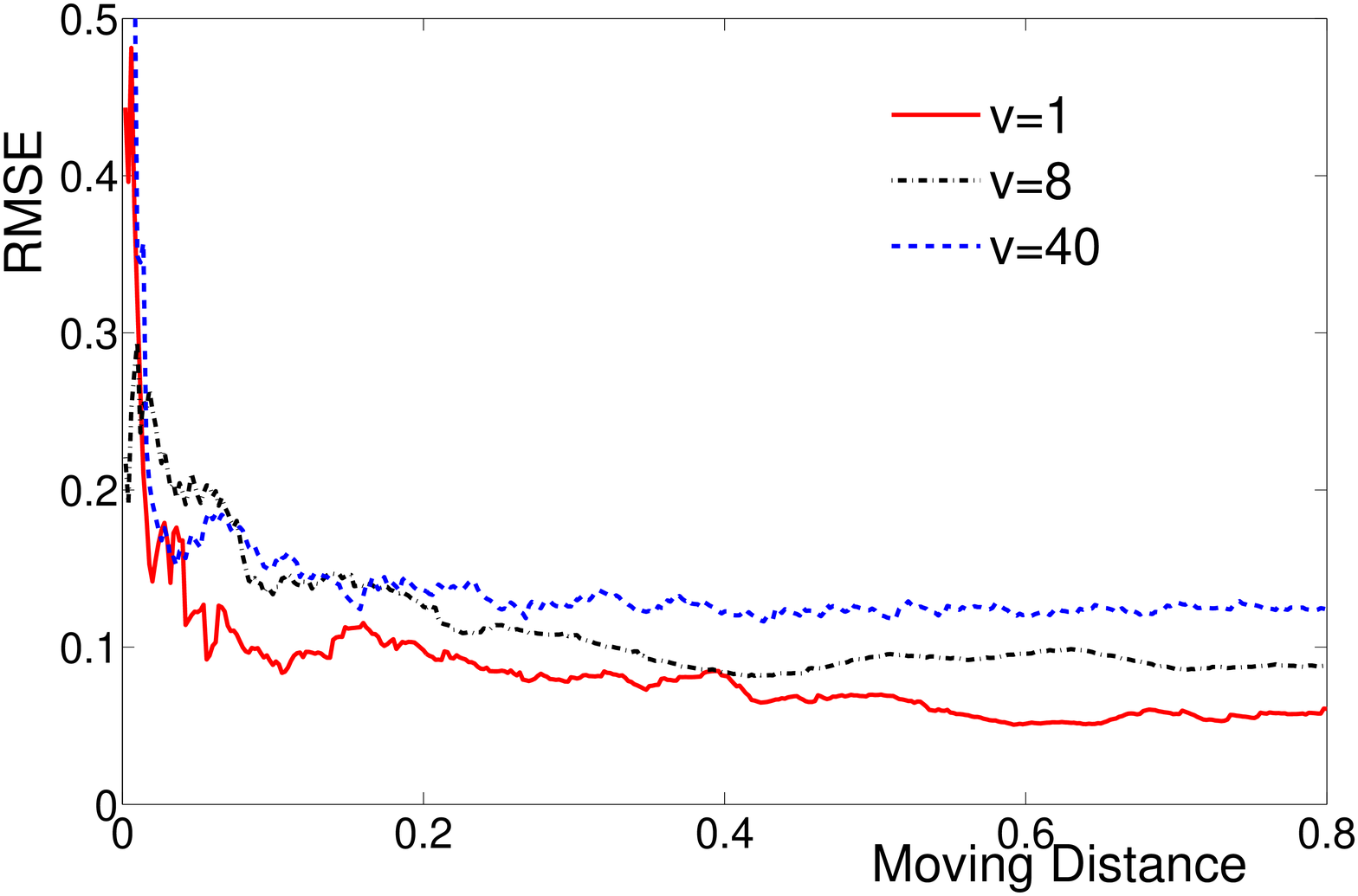}
\caption{ Comparison of RMSE as a function of boundary-moving distance (one side) at different boundary-moving speeds $v=1,8,40$. The other parameters are set as $b_0=0.5, a=0.2, L=1.0, \hbar=1.0$ and the initial state is the 100th eigenstate ($m=100$). $\tau$ is set to be $\tau=0.80, 0.10, 0.02$ to ensure the total moving distance equals 0.80 in all three experiments.}
\label{fig5}
\end{figure}

In Figs.~\ref{fig4} and \ref{fig5} we show the numerical results of RMSE for different values of $a$ and 
$v$. For the convenience of comparison, the horizontal axes in Figs.~\ref{fig4} and \ref{fig5} are chosen to 
be  the moving distance instead of the moment of time $t$. In other words, we compare the RMSE when the moving boundaries reach the same location. In Fig.~\ref{fig4}, the fixed parameters  are  $b_0=0.5, L=1.0, v=2.0, \tau=0.4$, and $a$ is chosen to be $a=0.1,0.2,0.3$. In Fig.~\ref{fig5}, $a$ is fixed at 0.2 while $v$ is set to be $v=1,8,40$, and accordingly the total moving times are $\tau=0.80, 0.10, 0.02$. 
All the other parameters in Fig.~\ref{fig5} are the same as those in Fig.~\ref{fig4}. In all these cases, $R$ is initially equal to zero, and increases very rapidly (the top is not visible in some figures), and then begins to decrease. We observe that $R$ decreases with oscillations in all the cases, and 
finally saturate at a finite value. The initial jump of RMSE from 0 to a large number is probably due to the local nature of the classical dynamics and the nonlocal nature of the quantum dynamics. In a short time, the classical transition probabilities cannot fully reflect the global feature of the system. 
However, even on a short time scale, the quantum transition probabilities can fully reflect the global feature of the system. Hence, at the initial stage of the driving process, these two transition probabilities differ substantially. Fig.~\ref{fig4} shows that the larger the parameter $a$ is, which means that the system becomes more chaotic, the more rapidly $R$ falls, and the smaller value $R$ saturates at. This result indicates that these two cumulative probabilities become closer when the system becomes more chaotic, and could be explained as follows: For a 2D system, the more chaotic it is, the better the quantum-classical correspondence principle applies \cite{Footnote}. Fig.~\ref{fig5} shows that as the boundary-moving speed increases, the two cumulative transition probabilities become more distinct. A similar result was obtained in the 1D piston system \cite{Quan2012}. This result can also be explained by the fact that the quantum transition probabilities can always reflect the global property of the system, while the classical transition probabilities cannot unless the boundaries move slowly and the classical particles collide frequently with the boundaries.

\begin{figure}[tbh]
\includegraphics[width=9cm, angle=0]{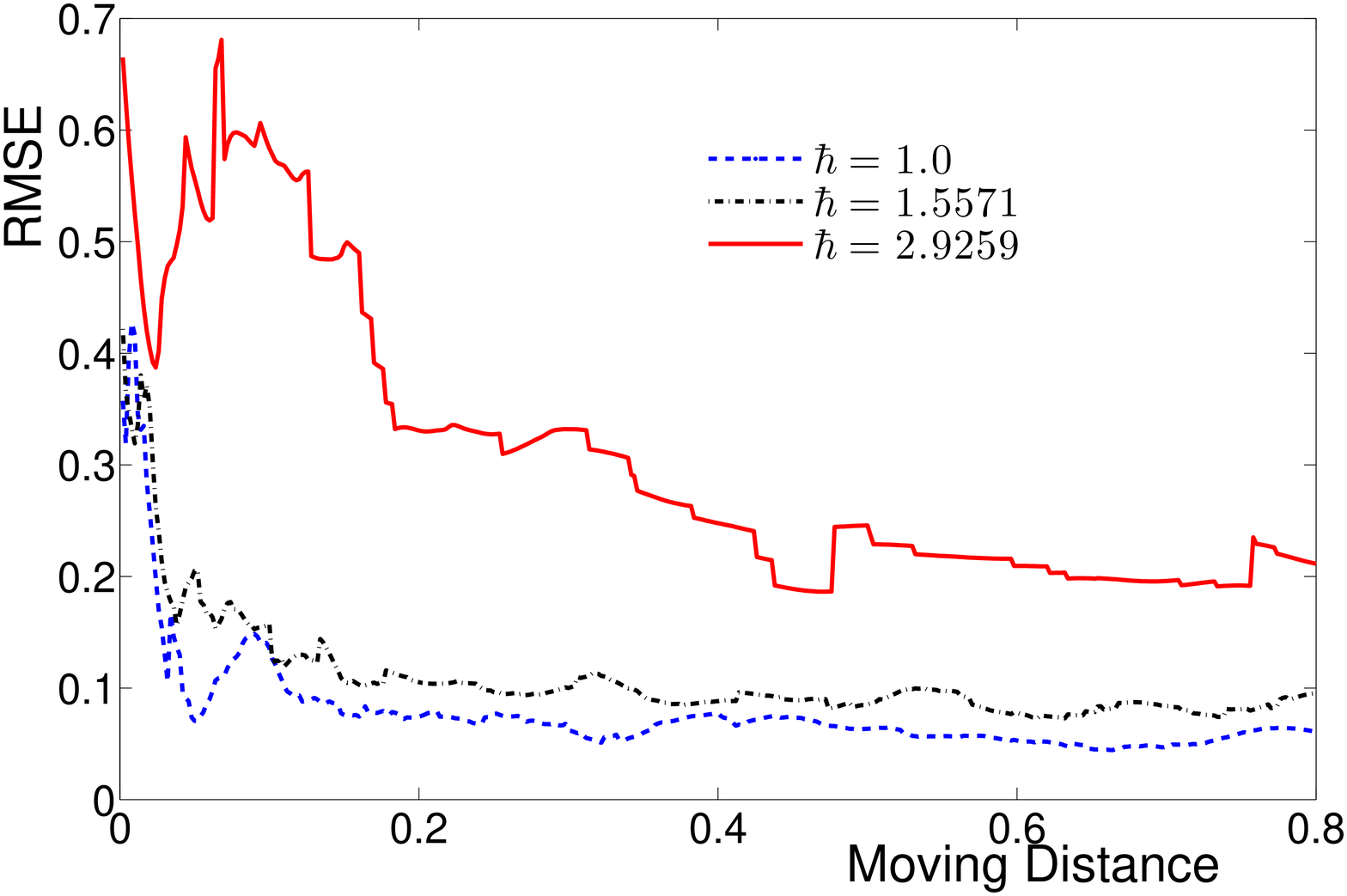}
\caption{ Comparison of RMSE as a function of boundary-moving distance (one side) for different values of $\hbar$. The other parameters are set as $b_0=0.5, a=0.2, L=1.0, v=2, \tau=0.4$. The initial energy of the system is set to be the 100th eigen energy when $\hbar=1.0$. The quantum number of the initial states are equal to $m=100$, $m=40$ and $m=10$ respectively.}
\label{fig6}
\end{figure}

We have also studied the effect of the Planck's constant $\hbar$ on the convergence of the quantum and the classical transition probabilities. As is known, quantum 
and classical predictions must agree when the Planck constant approaches zero ($\hbar \xrightarrow{} 0$) \cite{Bokulich2008}. Therefore, it is interesting to see how $R$ 
changes in this chaotic system when we adjust the value of the Planck constant $\hbar$. For this purpose, we should keep the energy, instead of the quantum number, of the initial state as a constant. Except that, all the other parameters $b_0$, $a$, $v$, $L$ and $\tau$ are fixed. As we cannot guarantee the accuracy of  numerical results when we decrease the value of $\hbar$ (see the Appendix), we increase the value of $\hbar$ and present the results in Fig.~\ref{fig6}. We clearly observe the increase of the saturated value of the RMSE when $\hbar$ increases. This result implies that the difference between these two cumulative transition probabilities becomes more prominent with the increase of $\hbar$. Although we cannot give accurate numerical results for a smaller value of $\hbar$ for computational reasons, our results in Fig.~\ref{fig6} serve as an indirect evidence that, in this chaotic system, the distance between these two transition probabilities will diminish when the value of $\hbar$ decreases. This is in accordance with the well-known correspondence principle that quantum mechanics is reduced to classical mechanics in the limit of $\hbar \xrightarrow{} 0$.

In this section, we have shown that under various conditions there does exist a quantum-classical correspondence principle of transition probabilities in a chaotic system. The more chaotic the system is, or the more slowly the boundaries move, the better the convergence between the quantum and the classical transition probabilities becomes. Also, we indirectly show that the smaller the value of the Planck constant $\hbar$ is, the better the convergence between these two cumulative transition probabilities is. Last but not least, we would like to mention that the correspondence principle between these two transition probabilities may break down in the long time limit (after the so-called Ehrenfest time) in a chaotic system \cite{Berry1979,Casati1986,Heller1993,Zurek2002}. But in the case of nonequilibrium driving, especially a transient driving process as is usually the case in the study of FT, the correspondence principle is still valid.

\section{\label{sec:level1}Conclusion}



In this article we numerically study the correspondence principle for work distributions in a prototype model of quantum chaos---a driven ripple billiard system. The quantum (or classical) work distribution function is determined by two factors: (1) the initial thermal equilibrium distribution $P_{A}^{Q}(m)$ (or $P_{A}^{C}(m)$) and (2) the transition probabilities $P^{Q}(n^{B}|m^{A})$ (or $P^{C}(n^{B}|m^{A})$). Since the initial distribution functions for the classical and the quantum cases 
are approximately equal,  the correspondence principle between work distributions is simplified to  the correspondence principle between transition probabilities. Unlike the 1D integrable systems \cite{Jarzynski2015}, we cannot employ analytical approaches due to the lack of the semiclassical (WKB) wave function in a fully chaotic system \cite{Berry1977b,Berry1979,Berry1983,Ott2002,Stone2005,Bokulich2008}.
Instead, we numerically calculate both the classical and the quantum transition probabilities. 
Compared with the 1D integrable systems, the transition probabilities in the chaotic system are less regular. 
In particular, the quantum transition probabilities are sparse and discrete, while the classical ones are diffusive and quasi-continuous. 
While these features make the correspondence principle in the chaotic system less ``evident", we still observe the convergence from the cumulative transition probabilities, thus demonstrate that the correspondence principle of the transition probabilities  applies in the ripple billiard system. 
Our numerical results indicate that the convergence, which is quantified by the statistical quantity RMSE, becomes better when the system is more chaotic or the driving speed gets slower. We also provide indirect evidences that the convergence becomes better when $\hbar$ decreases. 

We would like to emphasize that, similar to integrable systems \cite{Jarzynski2015}, this correspondence principle is a dynamic one (the quantum and classical transition probabilities converge when $\hbar \to 0$) instead of the usual static one (probability distributions in position space converge for large quantum number) \cite{Liboff1984}. Hence, in the context of semiclassical physics, our work complements extensive previous studies on the static correspondence principle. In the context of nonequilibrium quantum thermodynamics, our work complements the recent progress made in Ref. \cite{Jarzynski2015} and further justifies the definition of quantum work via two point energy measurements.

In the future it would be interesting to explore the dynamic correspondence principle in a quantum many-body system, where indistinguishability \cite{Gong2014} and the spin statistics effect will make the quantum-classical correspondence principle even more elusive. These problems are left for our future works.

\textbf{Acknowledgments} HTQ gratefully acknowledges support from the National Science Foundation of China under grants 11375012, 11534002, and The Recruitment Program of Global Youth Experts of China.
BW is supported by the National Basic Research Program of China (Grants No. 2013CB921903 and No. 2012CB921300) and the National Natural Science Foundation of China (Grants No. 11274024, No. 11334001, and No. 11429402).

\appendix
\section*{APPENDIX: NUMERICAL METHOD}
\renewcommand\theequation{A.\arabic{equation}}
\setcounter{equation}{0}
We consider the quantum and the classical dynamics of the ripple billiard system with both boundaries moving at the same speed $v$ and in the opposite directions. For the classical case, we numerically simulate the evolution of a classical particle undergoing Newtonian dynamics. Then we repeat the simulation while changing the initial location and the initial direction of the velocity of the particle. We make a histogram by counting the number of particles ($N_{in}$ in Eq.~(\ref{eq_pro})) which fall into the energy window $(E_n^B, E_{n+1}^B)$ at the moment of time $t=\tau$. This histogram gives the classical transition probability $P^C(n^B | m^A)$. 
In our numerical experiment, we repeat the simulation for 1 million times ($N_{total}$ in Eq.~(\ref{eq_pro})) for each set of parameters. We recall that the initial locations in the coordinate space and the directions in the momentum space are evenly sampled. 

For the quantum case, the evolution is described by the solution of the following time-dependent Schr\"{o}dinger equation:
\begin{equation}
-\frac{\hbar^2}{2m}(\partial_x^2+\partial_y^2)\psi(x,y,t)=i\hbar \partial_t \psi(x,y,t),
\label{eq_dyn}
\end{equation}
which is subjected to the boundary condition $\psi|_{\partial D}=0$, where
\begin{equation}
D=\{(x,y):-f(y,t)\leq x\leq f(y,t),0\leq y\leq L\},
\end{equation}
and
\begin{equation}
f(y,t)=b(t)-a\cos(2\pi y/L).
\end{equation}
In the following we set $L=1$ and $M=0.5$ for simplicity. One of us B. Wu and collaborators \cite{Wubiao2002} have given the solution to the Schr\"{o}dinger equation of the static ripple billiard system
\begin{equation}
-\hbar^2(\frac{\partial^2}{\partial x^2}+\frac{\partial^2}{\partial y^2})\psi(x,y)=E_n\psi(x,y),
\label{sch_sta}
\end{equation}
with the time-independent boundary
\begin{equation}
-f(y)\leq x\leq f(y),
\end{equation}
where 
\begin{equation}
f(y)=b-a\cos (2\pi y).
\end{equation}
To solve equation (\ref{sch_sta}), they suggest to straighten the boundaries by introducing a pair of curvilinear coordinates $(u,v)$
\begin{equation}
u=\frac{x}{2f(y)},v=y.
\end{equation}
In terms of coordinates $(u,v)$, the ripple billiard is transformed into a square billiard. Let us introduce a set of orthogonal and complete wave functions
\begin{equation}
\phi_{m,n}(x,y)=\sqrt{\frac{2}{f(y)}}\sin \left[m\pi \left(\frac{x}{2f(y)}+\frac{1}{2}\right)\right]\sin (n\pi y).
\label{phi}
\end{equation}
In the basis of $\phi_{m,n}(x,y)$, Eq.~(\ref{sch_sta}) can be transformed into a matrix equation
\begin{equation}
\sum_{m,n=1}^{\infty}H_{m'n'mn}B_{mn}^{l}=E_{l}B_{m'n'}^{l},
\end{equation}
and the hamiltonian matrix elements are
\begin{equation}
\begin{aligned}
H&_{m'n'mn}=\frac{m^2\pi^2\hbar^2}{4}\delta_{m',m}(I_{n'-n}^2-I_{n'+n}^2)\\
&+n^2\pi^2\hbar^2\delta_{m',m}\delta_{n',n}+n\pi^2\hbar^2a\delta_{m',m}J_{n',n}^2\\
&+a\pi^2\hbar^2\delta_{m',m}J_{n',n}^3-\frac{3}{2}a^2\pi^2\delta_{m',m}J_{n',n}^5\\
&+2mna\pi^3\hbar^2(K_{m'+m}^1+K_{m'-m}^1)J_{n',n}^2\\
&+2ma\pi^3\hbar^2(K_{m'+m}^1+K_{m'-m}^1)J_{n',n}^3\\
&-6ma^2\pi^3\hbar^2(K_{m'+m}^1+K_{m'-m}^1)J_{n',n}^5\\
&+2m^2a^2\pi^4\hbar^2(K_{m'-m}^2-K_{m'+m}^2)J_{n',n}^5,\\
\end{aligned}
\label{eq_H}
\end{equation}

where,
\begin{equation}
I_n^1=\left\{\begin{array}{ll}
0,&{n\rm \ is\ odd}\\
\frac{1}{\sqrt{b^2-a^2}}(\frac{b-\sqrt{b^2-a^2}}{a})^{n/2},&{n \rm \ is\  even}
\end{array}\right.
\label{i1}
\end{equation}

\begin{equation}
I_n^2=\left\{\begin{array}{ll}
0,&{n\rm \ is\ odd}\\
\frac{2b+n\sqrt{b^2-a^2}}{2(b^2-a^2)}I_n^1,&{n\rm \ is\  even}
\end{array}\right.
\end{equation}

\begin{equation}
K_n^1=\left\{\begin{array}{ll}
0,&{n=0}\\
-\frac{(-1)^n+1}{2n\pi},&{n\neq 0}
\end{array}\right.
\label {k1}
\end{equation}

\begin{equation}
K_n^2=\left\{\begin{array}{ll}
1/12,&{n=0}\\
\frac{(-1)^n+1}{n^2\pi^2},&{n\neq 0}
\end{array}\right.
\label {k2}
\end{equation}

\begin{equation}
J_{n',n}^2=I_{n'+n-2}^1+I_{n'-n-2}^1-I_{n'+n+2}^1-I_{n'-n+2}^1,
\end{equation}
\begin{equation}
J_{n',n}^3=I_{n'-n+2}^1+I_{n'-n-2}^1-I_{n'+n+2}^1-I_{n'+n-2}^1,
\end{equation}
\begin{equation}
\begin{split}
J_{n',n}^5&=I_{n'-n}^2-I_{n'+n}^2-\frac{1}{2}(I_{n'-n+4}^2+I_{n'-n-4}^2\\
&-I_{n'+n+4}^2-I_{n'+n-4}^2).
\end{split}
\end{equation}
For the ripple billiard with moving boundaries, we choose $\phi_{m,n}(x,y,t)$ as a set of orthonormal basis,
\begin{equation}
\phi_{m,n}(x,y,t)=\sqrt{\frac{2}{f(y,t)}}\sin\left[\frac{m\pi}{2}\left(\frac{x}{f(y,t)}+1\right)\right]\sin(n\pi y).
\label{phit}
\end{equation}
In comparison to Eq. (\ref{phi}), here we only replace $f(y)$ with $f(y,t)$. We expand the wave function of time $t$ $\psi(x,y,t)$ in the basis of $\phi_{m,n}(x,y,t)$
\begin{equation}
\psi(x,y,t)=\sum_{m,n}c_{mn}(t)\phi_{mn}(x,y,t).
\label{exp}
\end{equation}
Substituting $\psi(x,y,t)$ (\ref{exp}) into Eq.~(\ref{eq_dyn}), and taking the inner product with $\phi_{m'n'}(x,y,t)$, we obtain
\begin{equation}
\sum_{m,n}[H_{m'n'mn}(t)+i\hbar B_{m'n'mn}(t)]c_{mn}(t)=i\hbar\dot{c}_{m'n'}(t),
\label{final1}
\end{equation}
where $H_{m'n'mn}(t)$ is the $H_{m'n'mn}$ in Eq.~(\ref{eq_H}) with $b$ replaced by $b(t)$, and $B_{m'n'mn}(t)$ is
\begin{equation}
\begin{split}
B_{m'n'mn}(t)&=\left[\frac{1}{2}\delta_{m'm}+m\pi(K_{m'+m}^{1}+K_{m'-m}^{1})\right]\\
&\cdot(I_{n'-n}^1-I_{n'+n}^1)\dot{b}(t).
\end{split}
\end{equation}
Here $K_{m'\pm m}^1$ and $I_{n'\pm n}^1$ are defined in Eqs.~(\ref{i1}) and (\ref{k1}). We rewrite Eq.~(\ref{final1}) as
\begin{equation}
\dot{c}_{m'n'}(t)=\sum_{m,n}\left[\frac{1}{i\hbar}H_{m'n'mn}(t)+B_{m'n'mn}(t)\right]c_{mn}(t).
\label{sp1}
\end{equation}
This is an ordinary differential equation and we solve it numerically by using the Crank-Nicolson method, where 
Eq.~(\ref{sp1}) is approximated by
\begin{equation}
\begin{split}
&\frac{c_{m'n'}(t_{k+1})-c_{m'n'}(t_k)}{t_{k+1}-t_k}=\frac{1}{2}\sum_{m,n}\{\frac{1}{i\hbar}[H(t_{k})]_{m'n'mn}\\
&+B_{m'n'mn}(t_k)\}c_{mn}(t_k)+\frac{1}{2}\sum_{m,n}\{\frac{1}{i\hbar}[H(t_{k+1})]_{m'n'mn}\\
&+B_{m'n'mn}(t_{k+1})\}c_{mn}(t_{k+1}),
\end{split}
\label{final3}
\end{equation}
According to Eq.~(\ref{final3}), we can solve the coefficients of the $(k+1)$th timestep $c_{m'n'}(t_{k+1})$ from those of the $k$th timestep $c_{m'n'}(t_k)$. The initial coefficients $c_{mn}(0)$ can be easily calculated by decomposing the initial state in the basis of $\phi(x,y,0)$.

The biggest challenge in our numerical calculation is the computational resources. Notice that in Eq. (\ref{final3}), we need to multiply matrices and inverse the matrices in each timestep. If the size of the matrix is too large, the computational resources required will be unacceptably huge. Hence, properly choosing the cutoff of the matrix is the key point in the numerical calculation. In our calculation, the initial state is the 100th eigenstate ($m=100$), the total timestep is $\sim10^6$, and the cutoff dimension of the matrix is 2000 when $\hbar=1.0$. When we change the parameter $\hbar$, the quantum number of the initial state needs to be changed accordingly to keep the initial energy as a constant. When $\hbar$ decreases, the quantum number of the initial state changes into a larger number, so the size of the matrix must be chosen larger, which means the required computational resources will increase exponentially and this method will soon run out of computational resources. This is the reason why we cannot guarantee  the accuracy when we decrease the value of $\hbar$. This method may also fail when the moving speed is too fast or the quantum number of the initial state is too large, for that the cutoff dimension of the matrix 2000 is too small in these situations.

\bibliography{correspondence_principle}

\begin{thebibliography}{46}
\bibitem[{cha()}]{challenge}
\bibinfo{note}{As a general statement, one may say that whenever the classical
  equations of motion are integrable, e.g., rectangular or circular billiards,
  the quantum mechanical version of the billiard is completely solvable. When
  the classical system is chaotic, the quantum counterpart is generally not
  solvable, and there will be numerous difficulties in its quantization and
  evaluation.}
\bibitem[{Foo()}]{Footnote}
\bibinfo{note}{Please note that we do not mean the correspondence principle of
  transition probabilities is broken for integrable systems in 2D. What we mean
  is that for 2D integrable systems, the microcanonical ensemble that we sample
  in Fig. \ref{fig1} is no longer the proper counterpart of the $m$th
  eigenstate. Accordingly, the correspondence principle of transition
  probabilities in 2D integrable systems needs to be studied in a different
  setting.}
\end{thebibliography}
\end{CJK}
\end{document}